# Over four minutes relaxation of pyruvate using chemically and physically induced deceleration of relaxation

Josh P. Peters[a], Florin Teleanu[b,c], Huijing Zou[b], Ehtisham Rasool[a], Jan-Bernd Hövener[a], Alexej Jerschow[b], Andrey N. Pravdivtsev[a]†

[a] Section Biomedical Imaging (SBMI), Molecular Imaging North Competence Center (MOINCC), Department of Radiology and Neuroradiology, University Hospital Schleswig-Holstein, Kiel University, Am Botanischen Garten 14/18, 24118 Kiel, Germany. E-mail: andrey.pravdivtsev@rad.uni-kiel.de

[b] Department of Chemistry, New York University, New York, NY 10003, United States. E-mail: aj39@nyu.edu

[c] ELI-NP, "Horia Hulubei" National Institute for Physics and Nuclear Engineering, 30 Reactorului Street, Bucharest-Magurele, 077125, Ilfov, Romania

†Corresponding author: andrey.pravdivtsev@rad.uni-kiel.de

**Abstract**:

[1-$^{13}$C]pyruvate is the most widely used tracer for hyperpolarized metabolic magnetic resonance imaging, with profound applications in tumor and inflammation diagnosis as well as treatment monitoring. The most fundamental hurdle to broader application, however, remains the rapid polarization relaxation and the associated signal loss. Here, we report a method to address this challenge. Studying the nuclear spin relaxation dispersion of [1-$^{13}$C]pyruvate across magnetic fields from 8 µT to 9.4 T, as a function of additives, solvents, and preparation methods, allowed us to achieve relaxation times of up to four minutes. Such a long time could enable reliable quality control and nearly polarization loss-free transport, further boosting the power of hyperpolarized metabolic MRI.

**Main text**

The usefulness of hyperpolarized metabolic magnetic resonance[1,2] hinges on the relation of two periods of time: the time required for the agent to trace the desired process or function in vivo (metabolic activity), and the lifetime of the hyperpolarization (observable time). The first period is primarily determined by physiology, while the second period must account for time for quality control, sample transfer, administration, distribution in vivo, and tracing the desired function, while retaining a polarization level sufficient for detection.

Unfortunately, many long-lived tracers do not feature rapid metabolism or function. Some tracers, such as $^{15}$N-labeled molecules with chemically and magnetically isolated nitrogen, or certain $^{13}$C molecules[3], exhibit long relaxation times but do not undergo relevant metabolic transformations, limiting their utility for studying physiological processes like perfusion[4–6]. Others, such as $^{13}$C-labeled glucose, undergo rapid metabolism but suffer from rapid relaxation, impeding human application[7]. Some $^{15}$N molecules have a long relaxation time at high fields but a short one at low fields[8]. [1-$^{13}$C]pyruvate, however, features sufficiently fast metabolism and comparatively slow relaxation and thus has become the most widely used hyperpolarization tracer[1,9]. Even for pyruvate, however, many applications are limited by the signal-to-noise ratio, which affects spatial resolution and sensitivity.

The key factor for MRI is the polarization available at the time of detection. This factor is determined by (a) the initial polarization and (b) the loss between polarization and detection. Many efforts have been made to maximize the initial polarization, and to date, about 50% of the theoretical maximum has been achieved[10]. Reducing the loss between polarization and detection is equally important but has received less attention yet (e.g., by prolonging $T_1$[11] or by polarizing in the MRI[12,13]).

Here, we systematically investigated the sample composition of [1-$^{13}$C]pyruvate and identified conditions under which the relaxation time exceeds 4 minutes. We identified at least six factors to achieve this: chelation agent (EDTA), buffer (Tris), solvent deuteration ($D_2O$), sample degassing (no $O_2$), pyruvate deuteration, and the applied magnetic field (**Fig. 1**). Likely, the recently discovered effect of chemically induced deceleration of nuclear relaxation (CIDER) plays a key role in explaining the impact of Tris buffer[11].

To accelerate nuclear magnetic relaxation dispersion (NMRD) measurements, we first transferred polarization from fast-relaxing $^1$H (if not deuterated) to slower-relaxing [1-$^{13}$C] of pyruvate using INEPT[14,15] with refocusing at high field, yielding net magnetization on $^{13}$C. Then, the sample was transferred to the desired field, relaxed for a variable time, and then returned to the high field for observation. The magnetic field cycling (MFC) concept has previously been applied to measure NMRD[16] and estimate polarization losses for hyperpolarized pyruvate[15,17,18], however, detailed relaxation studies remain scarce. We assessed contributions from various interactions to the relaxation of [1-$^{13}$C]pyruvate by combining NMRD measurements under different conditions, molecular dynamics simulations, and ab initio computations. The ideal conditions for long-lasting $T_1$ are illustrated in **Fig. 1**.

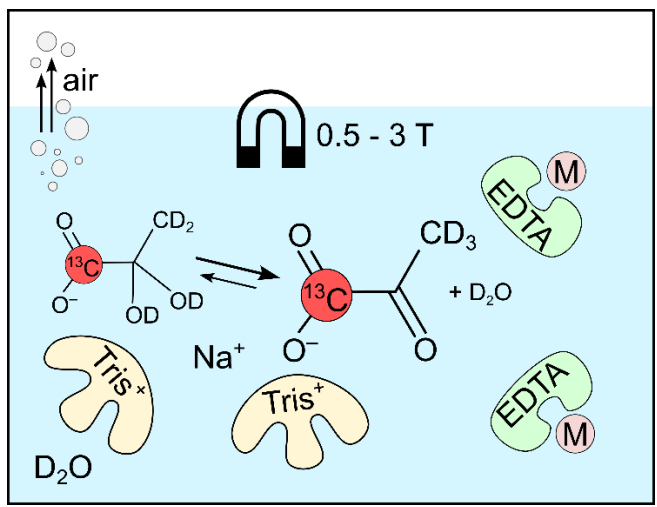

**Figure 1. Schematic view of [1-$^{13}$C]pyruvate and factors leading to more than 4 min 1-$^{13}$C relaxation time.** The relaxation of pyruvate is affected by intra- and intermolecular spin-spin interactions with nuclear spins and paramagnetic impurities such as $O_2$, chemical exchange, magnetic field, and others. Our investigations identified that the addition of Tris and EDTA, even in pure water, is beneficial for $T_1$. Degassing the sample, solvent deuteration, and the methyl group further prolong the relaxation time. EDTA chelates dissolved ions, like M = $Fe^{2+}$, $Mn^{2+}$. Tris could weakly coordinate with its positive charge to pyruvate, shielding it from exchange and impurities through the CIDER effect[11]. All experiments here were done at a neutral pH of 7.6±0.1. pKa of pyruvate is 2.5[19].

Key for this study was developing a highly automated MFC system[15], allowing us to systematically probe nuclear relaxation as a function of many well controlled parameters such as magnetic field and sample composition with reasonable efforts (**Fig. 2**). We would not have been able to acquire over 4300 free induction decay (FID) transients corresponding to a double of this amount of back and forth sample shuttling operations without the automation. This experimental series yielded nuclear magnetic relaxation dispersions at nine different conditions, as explained in **Fig. 2**. $T_1$ measurements at $B_0$ = 9.4 T were done without shuttling.

When pyruvic acid was dissolved in a double distilled water and adjusted with NaOH to neutral pH, 1-$^{13}$C relaxation was fastest at low fields ($\approx$ 8 µT – 8 mT, $T_1$ = 30.9 ± 0.9 s), slowest at intermediate fields ($\approx$ 0.1–1 T, $T_1$ = 48.6 ± 2.5 s), and accelerating again at high fields ($\approx$ 1 – 9.4 T, **Fig. 2#I**). The decay of $T_1$ at low field is associated with paramagnetic impurities and chemical exchange, while that at high magnetic fields is related to chemical shift anisotropy[20].

Substitution of $H_2O$ with $D_2O$ (**Fig. 2#H**) increased $T_1$ uniformly across the entire field range by $\approx$ 15 s. This effect is consistent with intermolecular dipolar relaxation being largely field-independent. The replacement of $^1H$ with $^2H$ reduced the water-induced dipole-dipole relaxation contribution by a factor of $(I_H(I_H+1))/(I_D(I_D+1))(\gamma_H/\gamma_D)^2 \approx 15.9$ (eq 11, Ref. [21]), where $I_H$ = ½ and $I_D$ = 1 are nuclear spin values and $\gamma$ corresponds to the gyromagnetic ratio of H and D. Nevertheless, sample **#H** still showed the shortest $T_1$ of all $D_2O$-containing samples at low fields (43.4 ± 1.5 s).

The addition of Tris buffer to $H_2O$, in conjunction with NaOH for pH adjustment, further increased $T_1$ across the entire field range, with the strongest effect observed at low fields (compare **Fig. 2#F** and **#I**). The lifetime was 50.0 ± 0.9 s at low field and reached 59.3 s in the intermediate field range at 2 T. We attribute the effect to Tris chelating paramagnetic species or coordinating with pyruvate via ionic interactions, thereby protecting it from rapidly relaxing species. This effect resembles the chemically induced deceleration of relaxation (CIDER) effect, in which additives shielded tracers from fast-relaxing and fast-exchanging nuclei such as protons[11].

To further investigate a possible chelating effect, we compared Tris-buffered solutions with and without EDTA (**Fig. 2#E** compared to **#F**). Adding EDTA increased $T_1$ only marginally (1–3 s), suggesting that the possible chelating property of Tris has a similar effect as EDTA here. Combining Tris and EDTA in $D_2O$ in **#D** resulted in a $T_1$ increase of $\approx$ 50 s at low fields and $\approx$ 20 s at high fields compared to **#I**, exceeding $\approx$ 115 s in the intermediate regime.

Interestingly, the $T_1$ of samples with Tris and EDTA was longer at low field than at high fields, with a maximum in the intermediate regime. This NMRD feature has practical value since relaxation at low and intermediate fields dominates hyperpolarization losses during transfer from the polarizer to the imaging system; prolonging $T_1$ in this range is therefore critical.

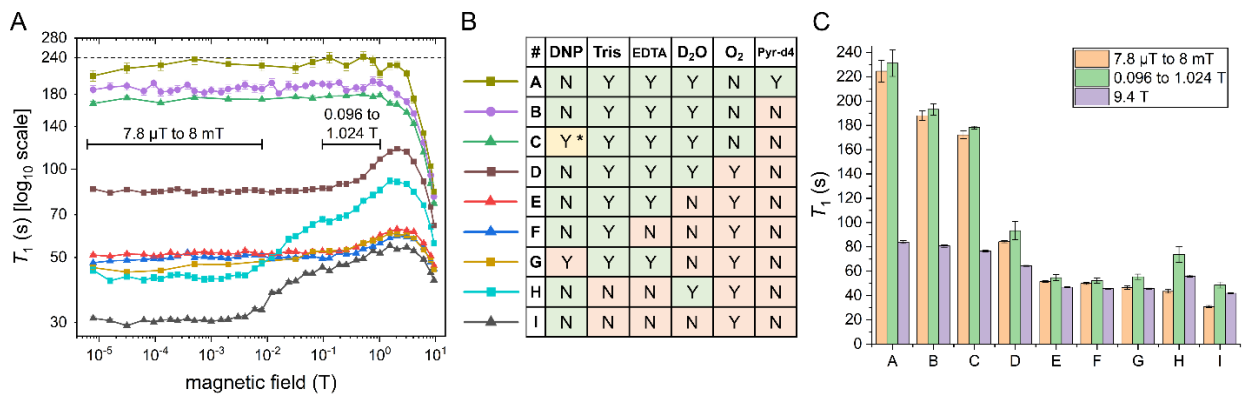

**Figure 2: NMRD of [1-$^{13}$C]pyruvate at fields from 7.8 µT to 9.4 T with different sample compositions.** Addition of Tris, and EDTA, the use of deuterated solvent and pyruvate, and sample degassing boosted $T_1$ of [1-$^{13}$C]pyruvate from about 30 to over 240 s at low magnetic fields (A) NMRD, (B) sample composition, (C) average $T_1$ values of these samples at three field regions. The field around 1 T gives the longest relaxation time across all samples, reaching a maximum of 242.1 ± 9.4 s at 0.512 T for the sample #A. Details: No INEPT was used on the **#A** due to the absence of protons in the pyruvate-d$_4$; therefore, fewer points have been sampled due to increased measurement time (>48 hours for the whole NMRD), while other NMRDs were measured with INEPT for acceleration of signal acquisition and signal enhancement. * Indicates that the radical in the dDNP sample has been filtered, and vitamin C was added after dDNP experiment.

Nonetheless, the decline in $T_1$ of **#D** from intermediate to low fields indicated contributions from either paramagnetic relaxation or chemical exchange. Degassing the samples via three freeze–thaw cycles further increased $T_1$, presumably by removing molecular oxygen with the triplet ground state, and eliminated the field-dependent dip at the low-to-intermediate field transition (**Fig. 2#B** compared to **#D**). At low field, $T_1$ increased from 84.2 ± 0.9 s (**#D**) to 188.0 ± 4.1 s (**#B**), indicating that oxygen was the dominant relaxation source at low field after all other sources had been eliminated.

Finally, deuteration of pyruvate itself extended $T_1$ at low and intermediate fields by an additional 20–30 s to more than 240 s (compare **Fig. 2#A** and **#B**). The average low field lifetimes were 224.5 ± 8.8 s for **#A** instead of 188.0 ± 4.1 s for **#B**. Pyruvate deuteration has a negligible effect in high and intermediate fields, but a sizeable 30 s in low fields in samples where other relaxation pathways were suppressed, raising the lifetime above 200 s for fields ≤3 T. Other studies observed only a marginal $T_1$ effect from deuteration[22] because experiments were performed at high fields, where relaxation is much faster and dominated by CSA.

Comparing these insights to the standard pyruvate sample used for metabolic imaging, we find that some conditions match, while others don't. For example, Tris and EDTA are included in the standard sample on an empirical basis, while NaCl is added for physiological isotonicity.

The $T_1$ dispersion of [1-$^{13}$C]pyruvate in a standard DNP sample showed a relatively fast relaxation at low fields (46.6 ± 1.4 s) and slower at intermediate fields (55.4 ± 2.2 s), followed by a steep acceleration of relaxation at high fields (**Fig. 2#G**). Still, this sample, although containing trityl radicals, has a longer relaxation time than the pure water sample **#I.**

When applying the measures suggested above to extend the lifetime (deuteration of the solvent, degassing, vitamin C, radical filtration), the lifetime increased significantly, for example, by more than threefold at low and intermediate fields up to 178.1 ± 1.2 s (**Fig. 2#C**).

Starting with 100% polarization, and assuming a period of 60 s between polarization and observation that should be sufficient for quality assurance and administration, and a magnetic field profile between dDNP system and NMR measured for our settings before (Ref. [15], **Fig. 6**, magnetic profile without transfer magnet), 27.4% will be retained for suboptimal (**#G**) and 70.4% for optimal (**#C**) samples. Hence, optimizing the sample composition could lead to a 2.6-times increase in signal-to-noise (SNR) ratio.

By altering the sample's chemical composition, we could isolate individual relaxation contributions and quantify them. This is achieved by subtracting NMRD profiles of sample pairs that differ in a single chemical feature. Then, we fitted the field dependence of these isolated interactions to a corresponding theoretical model and further compared them with predicted rates from molecular dynamics (MD) simulations, generating a detailed picture of the main factors limiting the $T_1$ (**Fig. 3**).

First, we highlight the presence of two primary peaks in $^{13}C$ spectrum corresponding to the oxo (Py) at ~170.3 ppm and hydrated (PyH) at ~178.7 ppm pyruvate forms with ~8.4 ppm chemical shift difference, which at neutral pH and at room temperature are in a ratio of $p_{Py}:p_{PyH} \sim 10:1$ (see SI)[19]. Chemical exchange between Py and PyH is illustrated on **Fig. 1**. Considering rapid exchange, 0.1 - 0.2 s$^{-1}$ (SI and Ref [23]), compared to longitudinal decay rates, >0.015 s$^{-1}$, the observed relaxation rate of the carboxylate $^{13}C$ in quantified Py is a weighted-average of two relaxation rates: $R_1^{obs} = p_{Py}R_1^{Py} + p_{PyH}R_1^{PyH}$ (See SI for mathematical derivation). This aspect is particularly important in the case of intra-molecular dipole-dipole coupling, where the additional hydroxyl protons enhance the relaxation rate of the PyH species and thus the observed $T_1$ of Py as well. At the same time, solvent deuteration will reduce intermolecular dipolar relaxation for both Py and PyH but will also decrease the intramolecular dipolar relaxation rate for PyH, as the hydroxyl protons are replaced with deuterium. Other relaxation interactions, such as CSA and paramagnetic relaxation enhancement (PRE), are assumed to be identical for the two species.

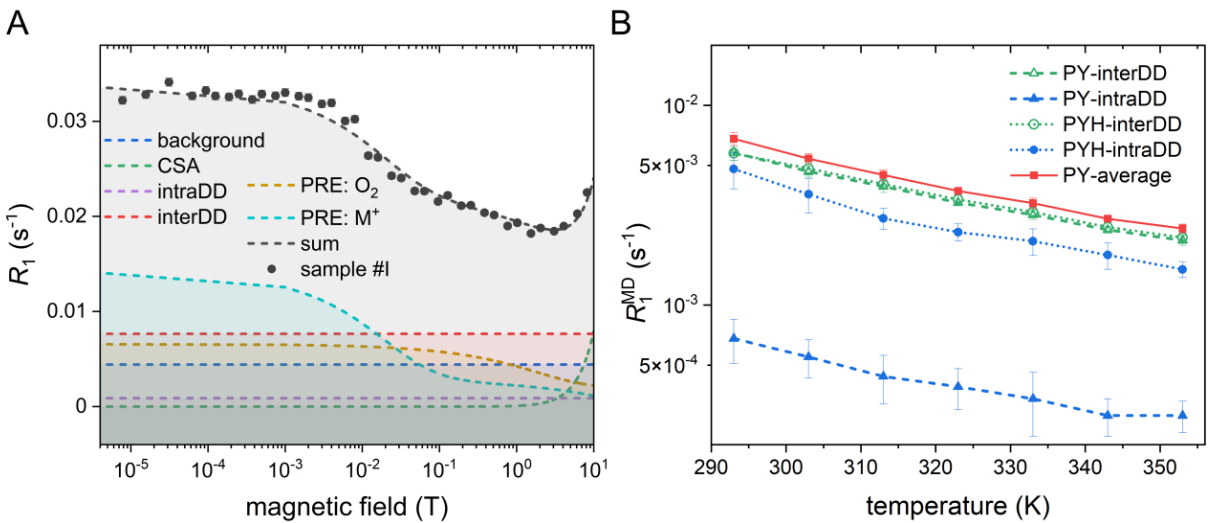

**Figure 3. Relaxation contributions to the relaxation of [1-$^{13}$C]pyruvate.** A) Field dependence of identified relaxation contributions (dashed lines) adding up to the NMRD profile of [1-$^{13}$C]pyruvate in water (black triangles), corresponding to sample **#I**. Several relaxation mechanisms —inter- and intramolecular dipole-dipole relaxation, PRE from $O_2$ and $M^+$, and CSA — have been isolated by subtracting pairs of NMRD profiles with different chemical compositions and fitting with theoretical equations, resulting in a fit (black line) closely matching the experimental data. B) Temperature dependence of predicted relaxation rates of longitudinal polarization due to $^1$H-$^{13}$C intra- (blue) or intermolecular (green) dipolar couplings for Py (triangles) and PyH (circles) species as extracted from molecular dynamics simulations for protonated pyruvate and solvent. Due to chemical exchange, the observable rate for Py was the population-weighted: $R_1^{\text{obs}} = p_{\text{Py}} R_1^{\text{Py}} + p_{\text{PyH}} R_1^{\text{PyH}}$ (red squares). The temperature dependence of the Py:PyH equilibrium was estimated from experimental measurements (See SI). Overall, dipolar relaxation decreases with increasing temperature. The maximum estimated lifetime of the molecule when only $^1$H-$^{13}$C dipole-dipole relaxation is considered is 2.44 ± 0.18 min at 293 K and 6.89 ± 0.35 min at 353 K.

Starting from the chemical composition that maximizes the $^{13}$C polarization lifetime (**Fig. 2#A**), we notice a dramatic increase in $T_1$ when the magnetic field is lowered from 9.4 T due to the decreasing contribution of the CSA, which scales quadratically with the external field (in the fast motion regime, $R_1^{\text{CSA}} = \frac{2}{15}(B_0 \cdot \Delta\sigma \cdot \gamma_{13C})^2 \tau_C^{\text{CSA}})^{24}$. This phenomenon impacts all displayed NMRDs, and at fields below 1 T, the CSA contribution is negligible. By fitting the high-field profile and using an MD-derived value for the rotational correlation time of Py at 293 K of $\tau_C^{\text{rot}}$=7.3 ps, we estimate $\Delta\sigma^{\text{exp}} \cong 136.5 \pm 1.6$ ppm in good agreement with our DFT simulations of Py in the gas phase ($\Delta\sigma^{\text{DFT}}$ = 133 ppm) (see SI).

The low field region of the slowest relaxing sample **Fig. 2#A,** we attributed to a "background" field-independent relaxation, $R_1^0$ = 0.0044 ± 0.0001 s$^{-1}$, resulting from the additive contributions of the intra- and inter-molecular $^2$H-$^{13}$C dipolar couplings and other remaining contributions from interactions like, possibly, spin-rotation. By running MD simulations of pyruvate in water, we estimated the population-weighted contributions from both $^2$H-$^{13}$C intra- and intermolecular dipole-dipole interactions (See SI), amounting to $R_1^{0,^2\text{H}-^{13}\text{C}}$ = 0.00043 ± 0.00003 s$^{-1}$, which represents only 10% of the observed background rate. When comparing deuterated **#A** and protonated **#B** pyruvate NMRD, we can determine the field-independent intra-molecular $^1$H-$^{13}$C dipolar contribution due to the distant methyl protons: $R_1^{\text{intraDD}}$ = 0.0008 ± 0.0002 s$^{-1}$ in good agreement with the MD-predicted value for Py system $R_1^{\text{Py-intraDD}}$ = 0.00068 ± 0.00017 s$^{-1}$.

Replacing the deuterated water with protonated water adds a field-independent inter-molecular $^1$H-$^{13}$C dipolar contribution to both Py and PyH and an additional intramolecular $^1$H-$^{13}$C dipolar contribution in PyH from hydroxyl protons: $R_1^{\text{inter\&intraDD}}(D_2O - H_2O)$ = 0.0076 ± 0.0002 s$^{-1}$ (comparing **2#E and 2#D**) and a similar value of 0.0095 ± 0.0006 s$^{-1}$ (comparing **2#I** and **2#H**). The individual MD-predicted dipolar coupling contributions are shown in **Fig. 3B** and amount to a population-weighted rate of 0.0062 ± 0.0005 s$^{-1}$, where only 6.5% comes from the intramolecular dipolar relaxation due to hydroxyl protons in PyH, such that solvent deuteration enhances $T_1$ by mostly reducing intermolecular interactions.

The dissolved molecular oxygen significantly reduces the $^{13}$C polarization lifetime (**2#D vs 2#B**) due to its paramagnetic relaxation contribution, $R_1^{\text{PRE}:O_2}$. The maximum value of $T_1$ is approximately 2 T, at an optimal balance between the decreasing CSA and the increasing PRE of O$_2$. The field-dependence of $R_1^{\text{PRE}:O_2}$ stems from the inter-molecular spectral density for which the tumbling regime is mostly dictated by the electron Larmor frequency and corresponding correlation time[24,25]. We estimate a correlation time of the dipole-dipole electron-carbon interaction of $\tau_C^{\text{PRE}:O_2}$ = 6.1±2 ps (given by oxygen's $T_{1,e}$) and an oxygen molar concentration between 100 – 200 µM by fitting $R_1^{\text{PRE}:O_2}$ (**SI**). These values are in good agreement with the literature[26].

The use of chelating agents (EDTA, Tris) revealed an additional prominent contribution from another paramagnetic species, which we believe might be transition-metal cation impurities, such as Mn$^{2+}$ or Fe$^{2+}$, at µM concentrations. The extra paramagnetic relaxation generates a second shoulder in the NMRD profiles around 0.1 T (**#H**, **#I**), indicating that the correlation time of the fluctuating intermolecular electron-carbon dipolar interaction is on the order of hundreds of picoseconds. This contribution is quenched by adding Tris or EDTA (or both), as they either chelate the paramagnetic ion M$^+$, such as Mn$^{2+}$ or Fe$^{2+}$, at µM concentrations or shield pyruvate via the CIDER effect[11]. In both cases, the distance to paramagnetic impurities increases, reducing the impact on the relaxation rate (**#I** vs **#F**). The low-field limits of these PRE contributions are estimated as $R_1^{\text{PRE}:M^+}$ = 0.0133 ± 0.0005 s$^{-1}$ (**#I** and **#E**) and $R_1^{\text{PRE}:M^+}$ = 0.0113 ± 0.0004 s$^{-1}$ (**#H** and **#D**).

Having identified relaxation contributions (**Fig. 3A, SI Section VI**), we can conclude that for the case of pyruvate in protonated water (sample **#I**), paramagnetic relaxation contributions become the most important sources of signal loss at low fields, while the high field is dominated by CSA and intermolecular $^1$H-$^{13}$C dipolar couplings.

Using MD simulations, we estimated how increasing temperature reduces both intra- and inter-molecular relaxation contributions, assuming full protonation (**Fig. 3B**). Likewise, the dipolar contributions due to deuterium coupling can be calculated by scaling the corresponding proton-induced rates with the factor $(I_H(I_H+1))/(I_D(I_D+1))(\gamma_H/\gamma_D)^2 \approx 15.9$ as mentioned previously. We estimate that the increase in temperature from 293 K to 353 K can lead to a 2.6-fold decrease in the intermolecular contribution and an averaged 2.8-fold decrease in the intramolecular dipole-dipole contribution. Thus, storing hyperpolarized $^{13}$C-labeled pyruvate at these temperatures in non-deuterated solvents is predicted to enhance T$_1$ by around 25% at 2 T. When, however, background relaxation sources are eliminated, and the system relaxes only through inter- and intramolecular dipole-dipole relaxation then the relaxation time of the fully deuterated system could be 39 min at 293 K and 110 min at 353 K. Probably at such a long relaxation times the other relaxation mechanisms would be dominated like paramagnetic impurities, partial protonation or even spin-rotation contribution. Unlikely these values could be reached. Still, it highlights the gap in knowledge regarding the relaxation contributions or the purity of the used samples.

We found that the relaxation of [1-$^{13}$C] pyruvate is affected by sample composition and preparation up to an order of magnitude (≈ 30 – 300 s), and up to 100% across magnetic fields (e.g., 50 s to 100 s or 240 to 70 s). Appropriate sample composition and preparation enabled us to extend the lifetime of [1-13C]pyruvate polarization to three minutes in a typical DNP sample and four minutes in an ideal sample. Previous studies did not observe this effect because experiments were mostly conducted at high fields, whereas relaxation is predominant and relevant for purification and transport at low fields. The effect was deciphered using simulations and can be used to significantly reduce polarization loss during transfer and quality assurance, e.g., at 1 T. The gain in polarization will further boost the power of hyperpolarized magnetic resonance to probe otherwise inaccessible processes. MD predicted dipole-dipole relaxation rates are lower than experimentally observed, tentatively indicating that even longer relaxation for pyruvate is possible.

Authors' emails, ORCIDs:
Josh Peters josh.peters@rad.uni-kiel.de, ORCID 0000-0003-1019-4067

Florin Teleanu ft2287@nyu.edu, ORCID 0000-0003-3845-0974

Huijing Zou hz2750@nyu.edu, ORCID 0009-0004-9274-7685

Ehtisham Rasool ehtisham.rasool@rad.uni-kiel.de, ORCID 0000-0002-9230-2094

Jan-Bernd Hövener jan.hoevener@rad.uni-kiel.de, ORCID 0000-0001-7255-7252

Alexej Jerschow aj39@nyu.edu, ORCID 0000-0003-1521-9219

Andrey N. Pravdivtsev, andrey.pravdivtsev@rad.uni-kiel.de, ORCID 0000-0002-8763-617X

## Author Contributions

A.N.P., J.-B.H.: conceptualization, J.P.: NMRD experiments, data analysis, E.R.: pyruvate chemical exchange study, F.T., H.Z., A.J.: relaxation simulations, J.P., F.T., A.N.P.: preparation of graphics, A.N.P., A.J., and J.B.H.: supervision, funding acquisition. All authors contributed to discussions and interpretation of the results, writing the original draft, and have approved the final version of the manuscript.

## ACKNOWLEDGMENT

We acknowledge funding from the German Federal Ministry of Education and Research (BMBF, 03WIR6208A hyperquant), DFG (555951950, 527469039, 469366436, HO-4602/2-2, HO-4602/3, EXC2167, FOR5042, TRR287). MOIN CC was founded with a grant from the European Regional Development Fund (ERDF) and the Zukunftsprogramm Wirtschaft of Schleswig-Holstein (Project no. 122-09-053). F.T. acknowledges funding from the European Union and Romania through the National Medical Project MySMIS SMIS Code 326475, and the Project ELI-RO/RDI/2024/14 SPARC funded by the Institute of Atomic Physics (Romania).

## ASSOCIATED CONTENT

### Supporting Information

Raw data, MD input parameters, DFT calculated geometries, and Mathematica scripts can be accessed via Zenodo DOI: https://doi.org/10.5281/zenodo.17175125.

Additional experimental data, fitting, relaxation analysis, results of MD simulations, and methods can be found in the supporting information (.pdf). The following references are cited in the supporting information Refs [26–34].